\documentclass[prl,twocolumn,showpacs,preprintnumbers]{revtex4-1}%
\usepackage{graphicx}
\usepackage{epstopdf}
\usepackage{array,multirow,amsmath,amsfonts}
\usepackage{color}
\usepackage{amsmath}
\usepackage{amsfonts}
\usepackage{amssymb}%
\providecommand{\U}[1]{\protect\rule{.1in}{.1in}}

\newcolumntype{Y}{>{\centering\arraybackslash}X}
\begin{document}
\preprint{ }
\title{Real-space charge distribution of cobalt ion and its relation with charge and spin states\\}
\author{Bongjae Kim}
\email{bongjae.kim@kunsan.ac.kr}
\affiliation{Department of Physics, Kunsan National University, Gunsan, 54150, Korea}
\date[Dated: ]{\today}

\begin{abstract}
 The charge state of an ion provides a simplified electronic picture of the bonding in compounds, and heuristically explains the basic electronic structure of a system. Despite its usefulness, the physical and chemical definition of a charge state is not a trivial one, and the essential idea of electron transfer is found to be not a realistic explanation. Here, we study the real-space charge distribution of a cobalt ion in its various charge and spin states, and examine the relation between the formal charge/spin states and the static charge distribution. Taking the prototypical cobalt oxides, La/SrCoO$_3$, and bulk Co metal, we confirm that no prominent static charge transfer exists for different charge states. However, we show that small variations exist in the integrated charges for different charge states, and these are compared to the various spin state cases.
\end{abstract}
\maketitle


\section{Introduction}
 The charge state is an important identity in chemical and materials physics. The concept of charge state (often, indistinguishably used as oxidation state) pedagogically explains the essence of the ionic bonding and provides a simple description of the bonding mechanism based on a few basic rules~\cite{Weller2014}. For example, oxygen is an element with strong electronegativity, and it tends to form a closed-shell configuration of $2s^2p^6$ with a formal oxidation charge state with 2-. When a multivalent transition metal, say, Mn, is combined with O, various combinations are possible. For each oxide of the form MnO$_2$, Mn$_2$O$_3$, and MnO, the Mn ion has an oxidation state of 4+, 3+, and 2+, respectively. Here, the idea of a formal charge state unambiguously explains the ionic bonding. For example, in the formation of MnO, two electrons of the Mn atom ($4s^23d^5$) are transferred to oxygen atom ($2s^2p^4$), O$^{2-}$ has a closed-shell configuration ($2s^2p^6$) and Mn$^{2+}$ will have five electrons in its $d$-shell ($d^5$). Likewise, for Mn$_2$O$_3$ and MnO$_2$, three and four electrons transfer from Mn to O, and each Mn ion will have a charge state of 3+ or 4+ with four or three electrons in $d$-shell ($d^4$ and $d^3$), respectively.

 However, the connection of a charge state to real atomic charge is not transparent. This is an old issue that has not been resolved, and related discussions are still actively ongoing~\cite{Raebiger2008,Jansen2008,Pickett2014,Karen2015,Walsh2018}. The basic idea of a charge state assumes static electric charges \emph{belong} to specific ions: Mn$^{3+}$ has four $d$-electrons and Mn$^{4+}$ three. However, when real-space integration of the static charges around the Mn atom is performed, the numbers of charges for different charge states are found to be almost the same~\cite{Luo2007}. This directly shows the failure of a simple ionic model, in which charge is transferred into or out of a specific atom.

 Many attempts to understand the nature of the oxidation states not based on direct counting or integration of electrons have been made. Notably, Raebiger \emph{et al.} considered the oxidation states of transition metals as different degrees of hybridization with neighboring ligands and explained the almost constant static electric charges of different charge states based on the negative feedback from transition metal - ligand bondings~\cite{Raebiger2008,Resta2008}. Based on this idea, Sit \emph{et al.} newly assessed the oxidation states with integer charge constrained~\cite{Sit2011}. In other approaches, such as the Bader charge analysis, the geometric distributions of the charges are often considered~\cite{Walsh2018}. Very recently, Pegolo \emph{et al.}, showed that the oxidation states can be defined as topological numbers~\cite{Pegolo2020,Grasselli2019} not from the electron density, where the idea of the modern theory of polarization is employed~\cite{Jiang2012,Resta2007}. Such discussions show the flaws in the conventional concept of oxidation states and suggest that the documented ionic radius is misleading despite its apparent usefulness in chemistry~\cite{Shannon1969}. This does not mean that the idea of charge status itself is a myth. For example, a clear distinction between different charge states exists based on the data from various experimental techniques, notably the spectroscopy. While the integrated charge inside the specific atom is not distinctive between different charge states, the distribution of the charges is, which is then reflected on the ground-state energy levels as can be spotted by using the spectroscopic technique~\cite{Raebiger2008}. Also, magnetic measurements can easily tell the difference between diverse charge states of transition metal ions, which is well-captured from first-principles density functional theory (DFT) calculations.

 While the connection of the charge state to a static electric charge is controversial, formal charge states are entities with chemical and physical meaning, and both from experiment and theory, clear identifications can be made. Thus, thinking that the difference in electron configurations among various charge states should be reflected in the real-space distributions of static charges, even though the distinction is not conspicuous, is natural~\cite{Luo2007}. Moreover, different phases with the same charge state, notably the spin states such as high- and low-spin, are also worth investigating from the view of the spatial charge distribution~\cite{Pickett2014}.

 In this study, employing DFT, we reexamine the relation of charge states and static distributions of the charges in transition metal ions. Taking Co ion, which is a multivalent ion with an additional degree of freedom in the spin states, we analyze the signatures in the static charge distributions of different phases. We chose LaCoO$_3$ and SrCoO$_3$ as test beds for different charge and spin states of the Co ions and compared the results with those from Co bulk systems.

 \section{Methods}
 All DFT calculations were carried out employing the Vienna ab-initio simulation package (VASP), which is the PAW approach\cite{Kresse1993,Kresse1996}. For the structure of LaCoO$_3$, Co$^{3+}$ case, we took a five-atom unit cell and employed a fully relaxed lattice parameter of 3.82 \AA. By replacing La with Sr with the same structure, we studied the Co$^{4+}$ case on an equal footing. We utilized the Perdew-Burke-Ernzerhof generalized gradient approximation as the exchange-correlation functional~\cite{Perdew1996}, and to treat the correlated Co-$d$ orbitals, we used Coulomb correlation of $U$=6.0 eV and $J$=0.9 in the form of DFT+$U$~\cite{Krapek2012,Park2020}. An energy cut-off of 500 eV was used, and for the $k$-point sampling, a Monkhorst-Pack grid of 10 $\times$ 10 $\times$ 10 was used. For bulk Co metal, we used a fully relaxed structure with $hcp$ symmetry. The same $U$ and $J$ values were employed with a mesh of 20 $\times$ 20 $\times$ 20 $k$-points.

 \section{Results and discussions}

\begin{figure}[t]
\begin{center}
\includegraphics[width=85mm]{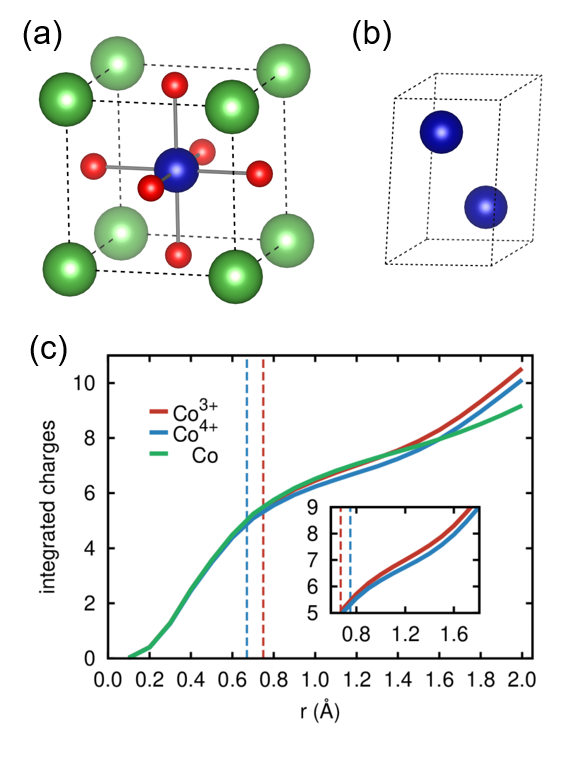}
\end{center}
\caption{(Color online) (a) Simulated crystal structure for Co$^{3+}$ and Co$^{4+}$. Co, O, and La/Sr atoms are represented by blue, red, and green balls, respectively. (b) The integrated $d$-electron density within a sphere around a Co atom is plotted as a function of the radius of the integration sphere for LCO (Co$^{3+}$), SCO (Co$^{4+}$), and bulk Co. The Shannon ionic radii for Co$^{3+}$ and Co$^{4+}$ (high-spin cases) are shown with dashed lines.}
\label{fig1}
\end{figure}

 Co is a typical multivalent ion, which can have 2+, 3+, and 4+ charge states in octahedral environments. Moreover, for the same charge state, notably for the Co$^{3+}$ ($d^6$) configuration, the ground-state energies between different spin states are reported to be small compared to those of other transition metal ions. This lead to multiple suggestions for the ground magnetic phases, especially for LaCoO$_3$ (LCO)~\cite{Rao2004}. To change the formal charge states with minimal extrinsic factors, we first take a cubic unit cell of LCO and simply switched La with Sr while keeping the same structure of five atom unit cell (See Fig.~\ref{fig1}(a)). For SrCoO$_3$ (SCO), the formal charge of Co is 4+, and the system is also discussed from the perspective of multiple spin states~\cite{Potze1995,Kunes2012}. For comparison, we chose bulk Co metal with a hcp structure (Fig.~\ref{fig1}(b)).

 Fig.~\ref{fig1}(c) shows the integrated electron density inside a sphere around a Co ion as a function of the radius. We can clearly see that the static charges associated with the Co ion are almost the same for Co$^{3+}$ (LCO) and Co$^{4+}$ (SCO). The radial accumulations of the charge have a similar shapes for both charge states. When compared to the neutral metal Co, the real space charge distribution is almost the same. The difference between the ion and neutral metal appears at large radius, but considering the Co-O bond length of about 1.91{\AA}, this is not due to the ionic charges. For bulk Co, the integrated charges are leaked for a larger radius of integration, meaning the electrons are itinerant. The Shannon ionic radii of Co$^{3+}$ and Co$^{4+}$ range from 0.67 to 0.75 {\AA}, so we can safely say that no actual charge transfer from the transition metal ion to oxygen ion occurs within the classical ionic radius, in agreement with the findings of previous studies~\cite{Luo2007,Raebiger2008}.

 On closer look, however, we observed that the Co$^{3+}$ and the Co$^{4+}$ curves start to bifurcate at a radius of about 0.7 {\AA} and that this eventually resulted in a sizable difference for larger integration spheres (see inset of Fig.~\ref{fig1}(b)). Here, the difference is much smaller than one electron, a formal value, and is noticeable outside the classical ionic radii. However a clear distinction between the Co$^{3+}$ and the Co$^{4+}$ charge distributions can be seen and the extent of the static charges is affected by the formal charge states. In general for transition metal ions, the ionic radius is larger for higher occupation (0.75/0.67 {\AA} for Co$^{3+}$/Co$^{4+}$), but, counterintuitively, we can see Co$^{4+}$ is more extended than Co$^{3+}$ from the real space distribution: at the same radius Co$^{3+}$ has more charges. The identical charge distribution around the ionic center directly disproves the variation in static electrons for different charge states; rather, as seen from the variations around the centers of Co and O ions, around \emph{r} = 0.95 \AA, this is more related to the character of the bonding. Note that due to the simple cubic assumption of the lattice, the magnetic moments of the Co ions in our calculations are 1.5, and 2.8 $\mu_{B}$ for LCO and SCO in their ground states, respectively, which are values in between the ideal high-spin and low-spin phases.

\begin{figure}[t!]
\begin{center}
\includegraphics[width=85mm]{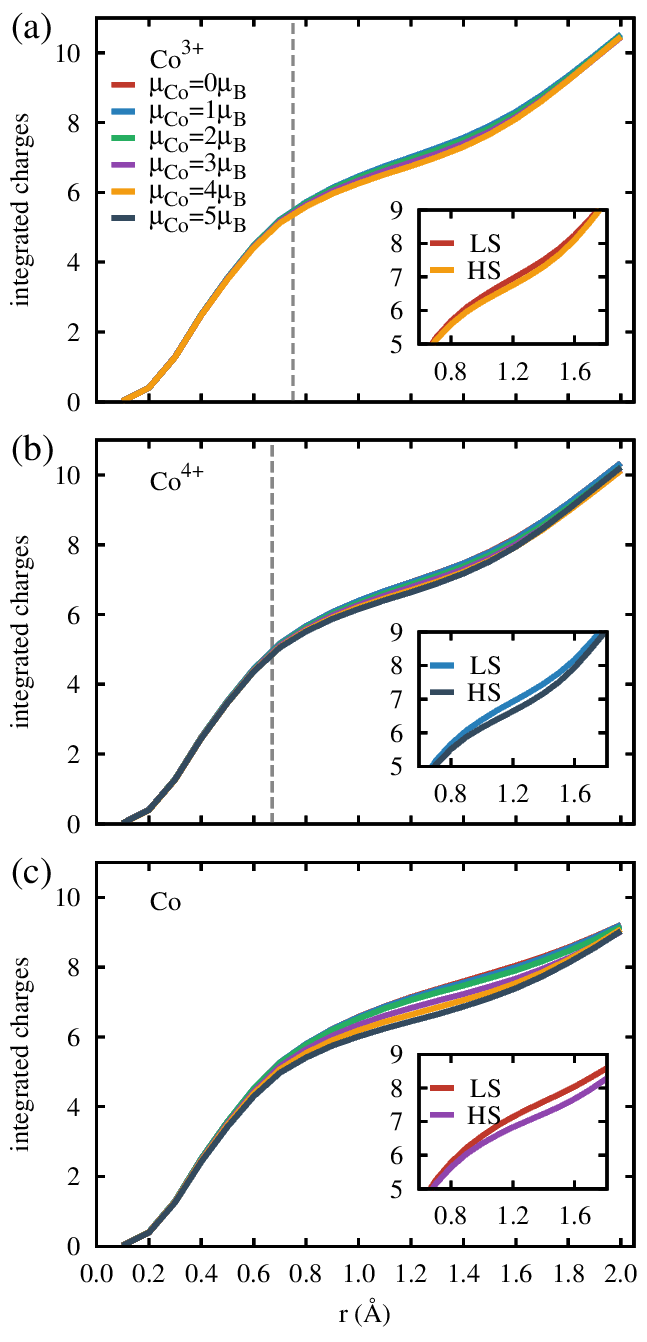}
\end{center}
\caption{(Color online) Integrated $d$-electron density within a sphere around a Co atom as a function of the radius of the integration sphere for (a) LCO, (b) SCO, and (c) bulk Co. The dashed lines are the Shannon ionic radii for each case. Various spin states are plotted. For each cases, blow-up graphs of the HS and the LS configurations, $t_{2g}^4e_g^2$ and $t_{2g}^6e_g^0$ for LCO, $t_{2g}^4e_g^1$ and $t_{2g}^5e_g^0$ for SCO, and S=0 and 3/2 for Co, are shown in the inset.}
\label{fig2}
\end{figure}

\begin{figure}[t!]
\begin{center}
\includegraphics[width=85mm]{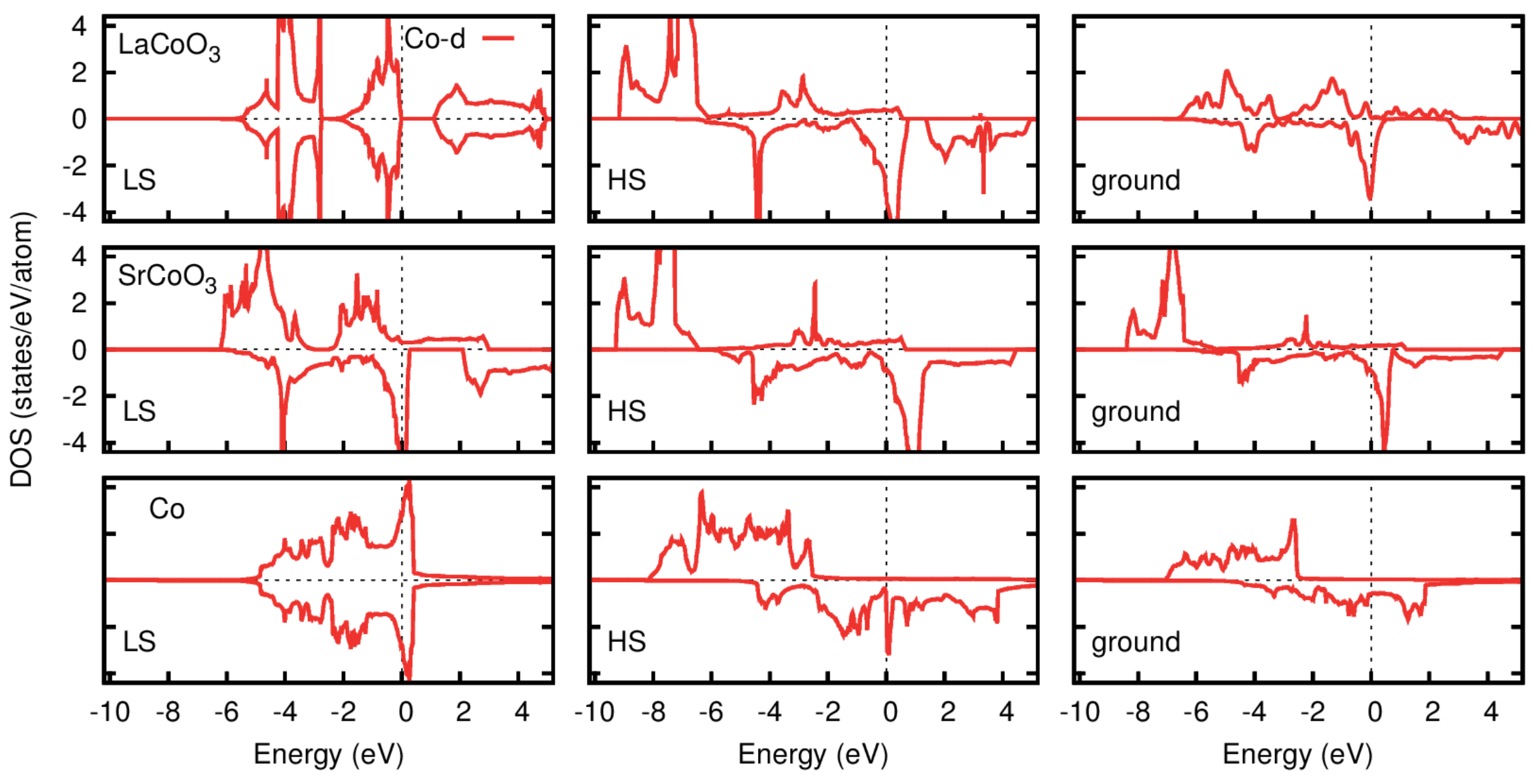}
\end{center}
\caption{(Color online) Co-$d$-projected partial density of states (pDOS) for various spin states: LS, HS, and ground states.}
\label{fig3}
\end{figure}

 Now, we discuss the different spin states: For LCO, three spin configurations were suggested for $d^6$ occupancy: high-spin (HS, $t_{2g}^4e_g^2$) with $S=2$ (4$\mu_B$), intermediate-spin (IS, $t_{2g}^5e_g^1$) with $S=1$ (2$\mu_B$), and low-spin (LS, $t_{2g}^6e_g^0$) with $S=0$ (0$\mu_B$). For SCO, $d^5$ occupancy of Co can form states from a HS of $S=5/2$ ($t_{2g}^3e_g^2$, 5$\mu_B$) to a LS of $S=1/2$($t_{2g}^5e_g^0$, 1$\mu_B$). By artificially controlling the difference between spin-up and -down occupied electrons, we have various spin configurations for each LCO and SCO case. The dependence of integrated static charges on the radius of the sphere are shown in Fig.~\ref{fig2}. The related partial density of states (pDOSs) of Co-$d$ orbitals for the HS and LS are shown in Fig.~\ref{fig3}.

 First of all, the various spin states show very similar behaviors of integrated charges as a function of the sphere's radius. As long as the formal charge states are the same for diverse spin configurations, the integrated charges should, in principle, be the same. Indeed, the radial charge shows more similarity among different spin states when compared to charge state variations (See Fig.~\ref{fig1}). Furthermore, despite the small variations between radii of 0.8 {\AA} and 1.6 {\AA}, the integrated charges eventually converge to the same value for larger volume, which is not the case for different charge states.

 Interestingly, from a radius of 0.8 {\AA}, the radial integrated charges show sizable divergences among different spin states for both LCO and SCO. For the same integration radius, the LS configuration has more charges, meaning more localized ions for LS than for the HS. In general, the specific spin states are determined from the competition between crystal field energy splitting between the $t_{2g}$ and the $e_g$ levels, $10Dq$, and Hund's exchange $J_H$. When the system is driven to the LS state, the effective $10Dq$ becomes larger, and the $e_g$ level moves to higher energy as shown in Fig.~\ref{fig3}: From the pDOS, we can see that the energy separation between the unoccupied $e_g$ and $t_{2g}$ orbitals is larger for the LS. This increase of $10Dq$ is equivalent to an effective decrease of the Co-O bond lengths, or volume, hence enhancing the localization.

 We also performed similar calculations for bulk metal Co, and we set the LS and the HS as S = 0 and 3/2, respectively. For bulk Co, the system possesses metallic bonds, where the electrons are highly itinerant compared to covalent and ionic bonds. Interestingly, we also observe a similar difference that arises for diverse magnetization. This shows that the relative occupation of different orbitals affects the character of bonding even for metallic bonds.

 While the distinctions of different charge and spin states are subtle from a static charge distribution, these can be well-captured using spectroscopic measurements. From the local cluster perspectiv, the different charges states of Co$^{3+}$ and Co$^{4+}$ are represented as $d^6+d^7\underbar{L}$ and $d^5+d^6\underbar{L}$, where $\underbar{L}$ is a ligand-hole state. This means the ground state configuration, which assumes a mixed state of Co-$d$ and O-$p$, cannot be captured with a Co-$d$ only perspective and that the spectroscopic distinction between the two states is not simply from a real space charge distribution. As explained by Raebiger \emph{et al.}, this ground charge state change can be viewed as a feedback effect, not the an actual charge transfer~\cite{Raebiger2008}. For the same charge state, the difference between the HS and the LS is primarily from the $10Dq$ parameter, which is in competition with Hund $J_H$, and as a result, changes the spectroscopic signal, from, such as, x-ray absorption~\cite{Chin2017,Hu2004}.

 As previously mentioned, the magnetic features can distinguish different charge states~\cite{Shen2005}. However, the measured magnetic responses are usually not only the contribution of the transition metal ion, but also those of its hybridization with nearby ions, O-$p$ in LCO and SCO. For instance, despite the formal magnetic moment of Co-$d$ being changed from 4$\mu_B$ to 5$\mu_B$ as HS Co$^{3+}$ is varied to HS Co$^{4+}$, the measured magnetic moment is not so distinguishable in either experiments or calculations. From the DFT calculation, we obtained the magnetic moments of the Co ion as 3.20 and 3.38$\mu_B$ for the HS LCO and SCO, respectively, and found that the other magnetic contributions were mainly from hybridized O-$p$ orbitals.

\begin{figure}[t!]
\begin{center}
\includegraphics[width=85mm]{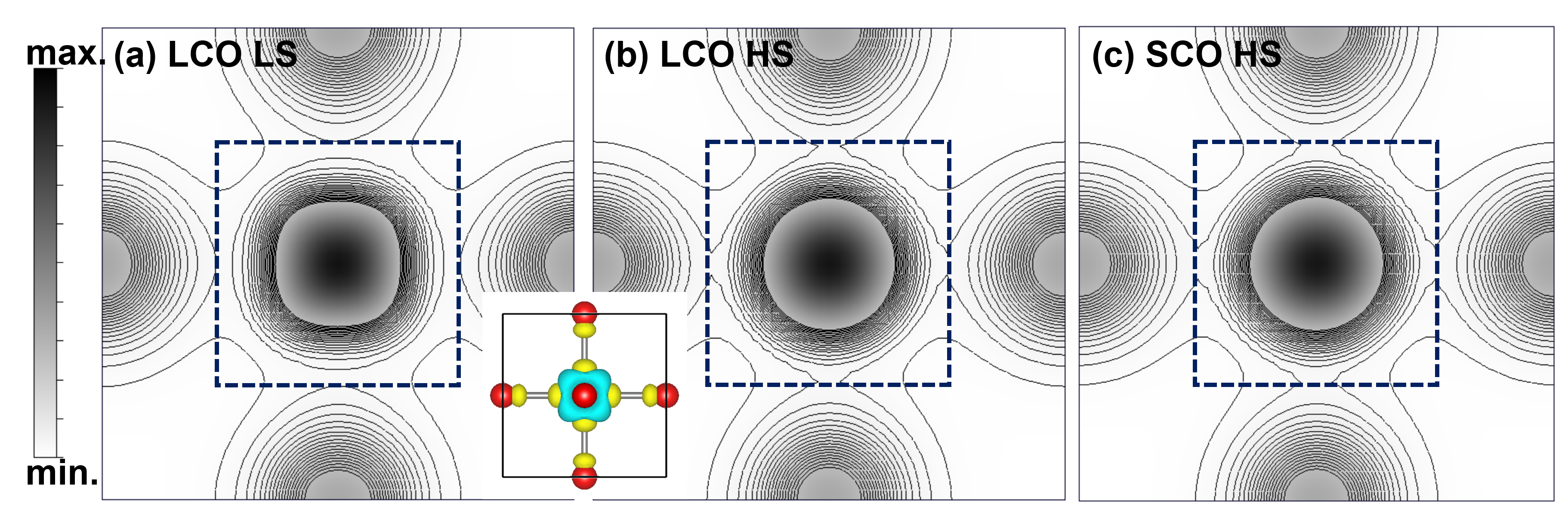}
\end{center}
\caption{Contour plots on a cross-sectional plane containing Co ions in the center, and O ions at the edge centers: charge densities for (a) LS LCO, (b) HS LCO, and (c) HS SCO cases. The dashed lines are guides for the eye.}
\label{fig4}
\end{figure}

 From the charge density plot, the detailed charge distribution, which cannot captured from the spherical integration, is found (see Fig.~\ref{fig4}). By comparing Figs.~\ref{fig4}(a) and (b), we can see the prominent $t_{2g}$ feature for the LS when compared to the almost isotropic charge distribution of the HS case. From the charge density difference, in the inset between Figs.~\ref{fig1}(a) and (b), clear spatial differences can be seen. While the static charge integrations are similar for both cases, the different bonding characters are well identified. For the HS cases of LCO and SCO, negligible differences are seen in the charge density plot. As shown in Fig.~\ref{fig1}(b), for different charge states, the difference in the static charge integration is not prominent around the ionic center, but is noticeable for larger integration radii above 1.0 {\AA}, a bond center. Together with the charge density plot, we can say that for different formal charge states, not many changes in the local ion occur and that no static charge transfer occurs.

\section{CONCLUSIONS}

 In summary, we have studied the real space charge distribution of various charge and spin states of the Co ion. Contrary to the common idea, almost no difference is seen between different charge states and the spherically integrated charges for both Co$^{3+}$ and Co$^{4+}$. This directly proves that the simple counting of the electrons around an ion is not a good measure of the oxidation states as previously reported~\cite{Raebiger2008,Pickett2014,Luo2007} and that a rigorous definition is required~\cite{Jiang2012,Pegolo2020}. However, the details of the real space charge distribution contain valuable information about the oxidation and the spin states. For different oxidation states, one can observe small variations farther away from the classical ionic region coming from the different bonding characters such as hybridization and covalency. Also, for different spin-states, slightly enhanced localization for LS case can be found, which is due to an effective increase in the crystal field compared to the HS case.

\begin{acknowledgments}
I thank Sooran Kim, Kyoo Kim, Beom Hyun Kim, K.-T. Ko, and Jeongwoo Kim, for fruitful discussions.
This work was supported by research funds from Kunsan National University.
\end{acknowledgments}


\begin{references}
\bibitem{Weller2014} M. Weller, T. Overton, J. Rourke, and F. Armstrong,
\emph{Inorganic Chemistry}
(Oxford University Press, Oxford, 2014).

\bibitem{Raebiger2008} H. Raebiger, S. Lany, and A. Zunger,
{Nature {\bf 453}, 763 (2008)}.

\bibitem{Jansen2008} M. Jansen and U. Wedig,
{Angew. Chem. Int. Ed {\bf 47}, 10026 (2008)}.

\bibitem{Pickett2014} W. E. Pickett, Y. Quan, and V. Pardo,
{J. Phys.:Condens. Matter {\bf 26}, 274203 (2014)}.

\bibitem{Karen2015} P. Karen,
{Angew. Chem. Int. Ed {\bf 54}, 4176 (2015)}.

\bibitem{Walsh2018} A. Walsh, A. A. Sokol, J. Buckeridge, D. O. Scanlon, and C. R. A. Catlow,
{Nature Mater. {\bf 17}, 958 (2018)}.

\bibitem{Luo2007} W. Luo, A. Franceschetti, M. Verela, J. Tao, S. J. Pennycook, and S. T. Pantelides,
{Phys. Rev. Lett. {\bf 99}, 036402 (2007)}.

\bibitem{Resta2008} R. Resta,
{Nature {\bf 453}, 735 (2008)}.

\bibitem{Sit2011} P. H.-L. Sit, R. Car, M. H. Cohen, and A. Selloni,
{Inorg. Chem {\bf 50}, 10259 (2011)}.

\bibitem{Pegolo2020} P. Pegolo, F. Grasselli, and S. Baroni,
{Phys. Rev. X {\bf 10}, 041031 (2020)}.

\bibitem{Grasselli2019} F. Grasselli and S. Baroni,
{Nature Phys. {\bf 15}, 967 (2019)}.

\bibitem{Jiang2012} L. Jiang, S. V. Levchenko, and A. M. Rappe,
{Phys. Rev. Lett. {\bf 108}, 166403 (2012)}.

\bibitem{Resta2007} R. Resta and D. Vanderbilt,
\emph{Theory of polarization: A modern approach,
{Physics of Ferroelectrics: A Modern Perspective (Springer, Berlin, Heidelberg, 2007)}, pp. 31–68}.

\bibitem{Shannon1969} R. D. Shannon, and C. T. Prewitt,
{Acta Crystallogr. B {\bf 25}, 925 (1969)}.

\bibitem{Kresse1993} G. Kresse and J. Hafner,
{Phys. Rev. B} \textbf{47}, 558 (1993).

\bibitem{Kresse1996} G. Kresse and J. Furthm\"{u}ller,
{Phys. Rev. B} \textbf{54}, 11169 (1996).

\bibitem{Perdew1996} J. P. Perdew, K. Burke, and M. Ernzerhof,
{Phys. Rev. Lett.} \textbf{77}, 3865 (1996).

\bibitem{Krapek2012} V. K\ifmmode \check{r}\else \v{r}\fi{}\'apek, P. Nov\'ak, J. Kune\ifmmode \check{s}\else \v{s}\fi{}, D. Novoselov, Dm. Korotin,and V. I. Anisimov,
{Phys. Rev. B {\bf 86}, 195104 (2012)}.

\bibitem{Park2020} Hyowon Park, R. Nanguneri, and A. T. Ngo,
{Phys. Rev. B {\bf 101}, 195125 (2020)}.

\bibitem{Rao2004} C. N. R. Rao, Md. Motin Seikh, and C. Narayana,
Top. Curr. Chem. {\bf 234}, 1 (2004).

\bibitem{Potze1995} R. H. Potze, G. A. Sawatzky, and M. Abbate,
{Phys. Rev. B {\bf 51}, 11501 (1995)}.

\bibitem{Kunes2012} J. Kune\ifmmode \check{s}\else \v{s}\fi{}, V. K\ifmmode \check{r}\else \v{r}\fi{}\'apek, N. Parragh, G. Sangiovanni, A. Toschi, and A. V. Kozhevnikov,
{Phys. Rev. Lett.} \textbf{109}, 117206 (2012).

\bibitem{Chin2017} Yi-Ying Chin, Hong-Ji Lin, Zhiwei Hu, Chang-Yang Kuo, Daria Mikhailova, Jenn-Min Lee, Shu-Chih Haw, Shin-An Chen, Walter Schnelle, Hirofumi Ishii, Nozomu Hiraoka, Yen-Fa Liao, Ku-Ding Tsuei, Arata Tanaka, Liu Hao Tjeng, Chien-Te Chen, and Jin-Ming Chen,
{Sci. Rep. {\bf 7}, 3656 (2017)}.

\bibitem{Hu2004} Z. Hu, Hua Wu, M. W. Haverkort, H. H. Hsieh, H. -J. Lin, T. Lorenz, J. Baier, A. Reichl, I. Bonn, C. Felser, A. Tanaka, C. T. Chen, and L. H. Tjeng,
{Phys. Rev. Lett.} \textbf{92}, 207402 (2004).

\bibitem{Shen2005} Xiong-Fei Shen, Yun-Shuang Ding, Jia Liu, Zhao-Hui Han, Joseph I. Budnick, William A. Hines, and Steven L. Suib,
{J. Am. Chem. Soc} \textbf{127}, 6166 (2005).
\end{references}
\end{document}